\documentclass[sigconf, nonacm]{acmart}
\usepackage{graphicx}
\usepackage[font=small,labelfont=normalfont, textfont=normalfont]{caption}
\usepackage{csquotes}
\usepackage{booktabs}
\usepackage{multirow}
\usepackage{makecell}
\usepackage{algorithm}
\usepackage{algpseudocode}
\usepackage{fancybox}
\newenvironment{summarybox}%
{\begin{center}\noindent\begin{Sbox}\begin{minipage}{0.96\columnwidth}}%
{\end{minipage}\end{Sbox}\fbox{\TheSbox}\end{center}}

\setcopyright{acmlicensed}
\copyrightyear{2025}
\acmYear{2025}
\acmConference[Conference acronym '25]{X}{Month from--to,
  2025}{X, Y}
\begin{document}

% \title{Exploring the Potential of Large Language Models for Detecting Tangled Code Changes}
% \title{Noise-Free Method-Level Bug Datasets: Using LLMs to Identify and Exclude Non-Bugfix Changes in Bug-Fix Commits}
% \title{LLM-Based Detection of Tangled Code Changes for Noise-Free Method-Level Bug Datasets}
\title{LLM-Based Detection of Tangled Code Changes for Higher-Quality Method-Level Bug Datasets}
% Can LLM can ...: an exploratory study
\author{Md Nahidul Islam Opu}
% \email{opumni@myumanitoba.ca}
\affiliation{%
  \institution{SQM Research Lab\\ Computer Science\\ University of Manitoba}
  \country{Winnipeg, Canada}
}

\author{Shaowei Wang}
\affiliation{%
  \institution{Mamba Lab\\ Computer Science\\ University of Manitoba}
  \country{Winnipeg, Canada}
}

\author{Shaiful Chowdhury}
\affiliation{%
  \institution{SQM Research Lab\\ Computer Science\\ University of Manitoba}
  \country{Winnipeg, Canada}
}
%\renewcommand{\shortauthors}{Opu et al.}

\begin{abstract}
\balance
Tangled code changes, commits that conflate unrelated modifications such as bug fixes, refactorings, and enhancements, introduce significant noise into bug datasets and adversely affect the performance of bug prediction models. Addressing this issue at a fine-grained, method-level granularity remains unexplored. This is critical to address, as recent bug prediction models, driven by practitioner demand, are increasingly focusing on finer granularity rather than traditional class- or file-level predictions. This study investigates the utility of Large Language Models (LLMs) for detecting tangled code changes by leveraging both commit messages and method-level code diffs. We formulate the problem as a binary classification task and evaluate multiple prompting strategies, including zero-shot, few-shot, and chain-of-thought prompting, using state-of-the-art proprietary LLMs such as GPT-5 and Gemini-2.0-Flash, and open-source models such as GPT-OSS-120B and CodeBERT. 

Our results demonstrate that combining commit messages with code diffs significantly enhances model performance, with the combined few-shot and chain-of-thought prompting achieving an F1-score of 0.883. Additionally, we explore machine learning models trained on LLM-generated embeddings, where a multi-layer perceptron classifier achieves superior performance (F1-score: 0.906, MCC: 0.807). Applying our approach to 49 open-source projects improves the distributional separability of code metrics between buggy and non-buggy methods, demonstrating the promise of LLMs for method-level commit untangling and potentially contributing to improving the accuracy of future bug prediction models.
\end{abstract}

%   <concept_id>10011007.10011006.10011066</concept_id>
%   <concept_desc>Software and its engineering~Version control systems</concept_desc>
%   <concept_id>10010147.10010257.10010293.10010294</concept_id>
%   <concept_desc>Computing methodologies~Neural networks</concept_desc>
%  % <concept>
%  %  <concept_id>10003752.10010070.10010071</concept_id>
%  %  <concept_desc>Theory of computation~Machine learning theory</concept_desc>
%  %  <concept_significance>100</concept_significance>
%  % </concept>
%   <concept_id>10011007.10011074.10011099.10011692</concept_id>
%   <concept_desc>Software and its engineering~Software evolution</concept_desc>

% \ccsdesc[500]{Software and its engineering~Version control systems}
% \ccsdesc[300]{Computing methodologies~Neural networks}
% % \ccsdesc[100]{Theory of computation~Machine learning theory}
% \ccsdesc[100]{Software and its engineering~Software evolution}



\maketitle

\section{Introduction}
\label{sec:introduction}

Software maintenance often exceeds the cost of initial development~\cite{borstler_role_2016}, with bug detection and resolution alone consuming 50–70\% of total development expenses~\cite{wang_bugpre_2023}. Undetected bugs can result in cascading failures, reduced developer productivity, and increased technical debt~\cite{noauthor_boeing_2020, noauthor_most_2024}. Consequently, the research community has extensively investigated bug prediction techniques to facilitate timely and targeted maintenance interventions~\cite{chowdhury_method-level_2024, pascarella_performance_2020, wattanakriengkrai_predicting_2022, giger_method-level_2012, rahman_relationships_2017}. Early bug prediction models focused on class- or file-level granularity~\cite{zimmermann_predicting_2007, alsolai_predicting_2018, basili_validation_1996, bell_does_2011, gil_correlation_2017, sun_boosting_2025, yin_line-level_2025}. However, these models are less helpful in practice, as only a small fraction of code in large units is typically defective~\cite{mo_exploratory_2022, pascarella_performance_2020}. Line-level models offer finer granularity but are prone to false positives due to coincidental line similarity~\cite{chowdhury_method-level_2024, servant_fuzzy_2017, grund_codeshovel_2021}. This has driven growing interest in method-level bug prediction~\cite{chowdhury_method-level_2024, mo_exploratory_2022, pascarella_performance_2020, Shippey, giger_method-level_2012, Hata:2012, Menzies:2007}, which aligns more closely with real-world debugging practices~\cite{grund_codeshovel_2021, chowdhury_method-level_2024}. However, method-level bug prediction faces challenges including evaluation bias and noisy data~\cite{pascarella_performance_2020, chowdhury_method-level_2024}, and progress is constrained by the scarcity of clean, labeled datasets~\cite{chowdhury_method-level_2024, pascarella_performance_2020}.

A key obstacle in building clean bug datasets is the presence of \textit{noise} in training data~\cite{bird_fair_2009}, often caused by \textit{tangled code changes}~\cite{chowdhury_method-level_2024, partachi_flexeme_2020, herzig_impact_2016}. While Version Control Systems like Git are intended for atomic commits, developers frequently combine unrelated tasks, such as bug fixes, refactorings, and enhancements, into a single commit~\cite{herzig_impact_2013, dias_untangling_2015, kirinuki_hey_2014}, violating best practices and introducing tangled changes. Researchers typically use commit messages to identify bug-fix commits, assuming all associated modifications are relevant. However, this assumption fails in the presence of tangled commits, leading to false positives in labels, which degrades dataset quality and impairs bug prediction performance~\cite{herzig_impact_2016, herbold_fine-grained_2022}.

To mitigate the impact of tangled changes, several prior works have proposed various untangling techniques ranging from heuristic-based program slicing~\cite{muylaert_research_2018}, lexical and data-flow based models~\cite{partachi_flexeme_2020}, and graph-based learning using dependency and name flows~\cite{li_utango_2022,xu_detecting_2025}. While effective at the file or statement level, none of these methods have addressed the problem at the method-level granularity, leaving a gap that presents a unique opportunity for exploration. Method-level untangling cannot be achieved using file-level approaches, as a single file may contain multiple methods, only some of which are bug-related. Labeling all methods in a file as buggy is inappropriate and would produce numerous false positives. On the other hand, although statements can be aggregated to form method-level representations, determining whether an individual statement contributes to a bug fix requires understanding its broader context, including surrounding code, method semantics, and commit intent~\cite{tufano2019empirical, sintaha2023katana}. For developers and models alike, the relevant context may span from within the method to the entire file or even across files~\cite{sintaha2023katana}. For instance, in the \textit{refill} function of \textit{FixedIntervalRateLimiter.java} from \textit{Apache HBase Commit 84a5039} \footnote{\url{https://github.com/apache/hbase/commit/84a50393ee56d09abb68f54b44b64f5279bd33c9}}, the function signature modification appears to be a bug fix when viewed in isolation, as removing an argument typically alters the function’s logic; however, examining the full method reveals it is actually a refactoring change. Thus, simply applying a statement-level model repeatedly at the method-level would require redundant context analysis for each modified line, introducing unnecessary computational overhead. Therefore, distinct representations at different granularity are needed to effectively model bug-fix behavior while maintaining contextual relevance.

Given that, this study focuses on detecting tangled changes at the method-level by framing the task as a binary classification problem: determining whether a method-level code change is related to a bug fix (\textit{Buggy}) or not (\textit{NotBuggy}) based on commit messages and code diffs. To tackle this, we explore the use of \textit{Large Language Models (LLMs)}, which have demonstrated strong capabilities in analyzing natural language and source code~\cite{chew_llm-assisted_2023, nam_using_2024}. In contrast to prior approaches in code analysis, classification, and untangling that rely on language-specific algorithms~\cite{herzig_impact_2016, guo_interactively_2017} or complex models requiring extensive training and feature engineering~\cite{zeng2024colare}, LLM-based methods enable faster system development through their language-agnostic design. LLMs also provide advantages such as promptability, flexibility in input handling, and the ability to leverage contextual signals from both code and commit metadata. 

% We evaluate several prompting strategies, including zero-shot, few-shot, and chain-of-thought prompting, using both proprietary and open-source LLMs. Additionally, we develop machine learning models leveraging embeddings generated by LLMs. Finally, we evaluate the impact of the LLM-based technique by measuring improvements in the distributional separability of code metrics between \textit{Buggy} and \textit{NotBuggy} methods through statistical analysis. In general, the contribution of the paper is centered around four research questions.

We evaluate several prompting strategies using both proprietary and open-source LLMs. Additionally, we develop machine learning models using LLM generated embeddings. Finally, we evaluate the impact of the LLM-based technique by measuring improvements in the distributional separability of code metrics between \textit{Buggy} and \textit{NotBuggy} methods through statistical analysis. In general, the contribution of the paper is centered around four research questions.

\textbf{RQ1: \textit{Can zero-shot LLMs detect tangled code changes by using code diff and commit message?}}

We designed a classification-based approach that uses LLMs to detect tangled changes from method-level code diffs, using zero-shot prompting, with or without the corresponding commit messages. Through experiments on our curated gold dataset, we show that including the commit message significantly improves the performance across multiple LLM variants, achieving the highest F1-score of 0.879 using \textit{gpt-5}.

\textbf{RQ2: \textit{How do different prompting techniques influence the effectiveness of detecting tangled code changes?}}

We investigate the impact of advanced prompting strategies, such as few-shot, chain-of-thought, and a hybrid combination. Our analysis shows that LLMs can reason more accurately about change semantics when guided with structured prompts and examples. Among these, chain-of-thought + few-shot prompting with \textit{gpt-4o} achieves the most balanced performance (F1-score: 0.883) and demonstrates superior capability in handling complex inputs.

\textbf{RQ3: \textit{How well can embedding-based machine learning models detect tangled code changes?}}

We utilized LLMs to generate embeddings from commit messages and code diffs, which served as input for embedding-based classifiers. One of the models exhibits strong predictive capabilities, outperforming the results obtained in RQ1 and RQ2. Specifically, a Multi-layer Perceptron classifier achieved the highest F1-score of 0.906, demonstrating the effectiveness of embedding-based representations for detecting tangled changes.

\textbf{RQ4: \textit{What is the potential impact of LLM-based untangling on future method-level bug prediction models?}}

Using a \textit{Less-Noisy} dataset, created by filtering out noisy samples through our LLM-based approach, we observe that the distributional differences in various code metrics between \textit{Buggy} and \textit{NotBuggy} methods are significantly more than in the original noisy dataset. This finding shows promise for future machine learning (ML) models for bug prediction, as many of them rely on these code metrics to predict whether a method is bug-prone or not.

To enable replication, we share our data and code publicly.\footnote{\url{https://github.com/SQMLab/Tangled}}

\section{Related Works} 
\label{sec:relatedWorks}

This section reviews prior work on tangled code changes, their prevalence, impact, and untangling techniques, as well as LLM-based approaches for analyzing and classifying text and code. Together, these studies motivate our research.

\subsection{Tangled Changes}
Herzig and Zeller~\cite{herzig_impact_2013} first used the term \textit{Tangled Changes} and reported that up to 15\% of bug-fixing commits across several open-source Java projects contained tangled changes.
Subsequently, Tao and Kim~\cite{tao_partitioning_2015} later observed tangling in up to 29\% of revisions. Herbold et al.~\cite{herbold_fine-grained_2022} confirmed its prevalence and impact, showing that only 22–38\% of lines in bug-fix commits actually fix bugs. Kochhar et al.~\cite{kochhar_potential_2014} further found that 28\% of files in bug-fix commits contain no bugs. 

The prevalence of tangled changes greatly affects the historical analysis of source code. By labeling all the modified methods in a bug-fixing commit as buggy methods, it harms the bug prediction models significantly~\cite{herzig_impact_2016, chowdhury_method-level_2024}. Consequently, researchers focused on detecting and untangling commits~\cite{herzig_impact_2013, tao_partitioning_2015, guo_interactively_2017, muylaert_research_2018, partachi_flexeme_2020}.

Early techniques relied on static analysis and heuristics. Herzig et al.~\cite{herzig_impact_2013} introduced confidence voters based on file distance, change coupling, call graphs, and data dependencies to estimate task co-membership. Tao and Kim~\cite{tao_partitioning_2015} applied program slicing and pattern matching to cluster semantically related edits, achieving 69\% agreement with manual decompositions. Guo and Song~\cite{guo_interactively_2017} proposed CHGCUTTER, an interactive method using control and data dependencies to decompose composite changes while preserving syntactic correctness. Muylaert and De Roover~\cite{muylaert_research_2018} applied program slicing on abstract syntax trees to untangle composite commits by grouping fine-grained code changes according to their dependence graph slices. Pârtachi et al.~\cite{partachi_flexeme_2020} introduced Flexeme, which overlays name flows on program dependency graphs using a $\delta$-NFG structure to better capture lexical cues. Using Agglomerative clustering for commit untangling they developed a tool Heddle, which achieved improved clustering accuracy (F1-score 0.81) and runtime efficiency.

Graph-based modeling has emerged as a powerful approach for capturing complex code relationships. Shen et al.~\cite{shen_smartcommit_2021} proposed SmartCommit, representing diff hunks as graph nodes enriched with semantic links (hard, soft, refactoring, cosmetic), and achieved a decomposition accuracy between 71-84\%. Chen et al.~\cite{chen_untangling_2022} and Xu et al.~\cite{xu_detecting_2025} enhanced this approach by incorporating node attributes such as token content and control/data flow, achieving 7.8\% and 8.2\% higher accuracy than Flexeme, respectively.

Fan et al.~\cite{fan_detect_2024} introduced a Heterogeneous Directed Graph Neural Network designed to capture semantic dependencies without relying on explicit code links. By leveraging hierarchical graphs at the entity and statement levels, they achieved substantial untangling improvements, 25\% for C\# and 19.2\% for Java, outperforming Flexeme and SmartCommit without sacrificing time efficiency.

Recent work increasingly applies machine learning to learn untangling patterns from data. UTANGO~\cite{li_utango_2022} uses a Graph Convolutional Network to learn contextual embeddings of code changes, incorporating cloned code and surrounding context. Framed as supervised clustering, it outperforms Flexeme by 9.9\% accuracy. Dias et al.~\cite{dias_untangling_2015} and Liu et al.~\cite{liu_linking_2018} explored fine-grained tracking and clustering based on developer interactions and commit metadata to achieve high temporal and contextual resolution.

Although these methods have shown success at the file- or statement-level, the method-level remains unexplored, despite being preferred by practitioners and researchers in order to create a clean and noise-free method-level bug dataset~\cite{pascarella_performance_2020, chowdhury_method-level_2024}. Motivated by this gap, we investigate whether tangled changes can be detected from method-level diffs and their corresponding commit messages.

\subsection{LLMs for Code Analysis \& Classification}
LLMs have shown strong performance in natural language understanding and source code analysis~\cite{brown_language_2020, zheng2024few}. Trained on large corpora of text and code, they generalize well to unseen inputs and often perform effectively in zero-shot settings~\cite{kojima_large_nodate}. This versatility has fueled growing interest in using LLMs for text classification and code analysis~\cite{ahmad_unified_2021, feng2020codebert}. Building on these capabilities, our study explores LLMs for detecting tangled code changes by jointly analyzing commit messages and code diffs.

Prompt design is critical to optimizing LLM performance. Recent advances in prompt engineering have introduced zero-shot, few-shot, and chain-of-thought prompting~\cite{wei_finetuned_2022, brown_language_2020, wei_chain--thought_2023}. Zero-shot learning uses only task instructions, while few-shot prompting includes exemplar input-output pairs to guide responses, yielding improved performance across tasks~\cite{brown_language_2020}. Chain-of-thought prompting enhances reasoning by encouraging step-wise problem decomposition~\cite{wei_chain--thought_2023}. Motivated by their empirical success, we systematically evaluate these techniques for detecting tangled code changes.

LLMs are built on Transformer architectures, consisting of encoder and decoder modules~\cite{vaswani_attention_2017}. Encoders convert input text into high-dimensional embeddings that capture semantic relationships, making them effective for downstream classification~\cite{petukhova_text_2024, keraghel_beyond_2024}. We extend this utility to software engineering by generating embeddings from commit messages and code diffs, and training machine learning models to classify them as tangled or untangled.

LLM embeddings are also used in commit classification tasks~\cite{zeng2025first, zeng2024colare}, which classifies commits into maintenance activity types. CodeBERT~\cite{feng2020codebert} has been used in COLARE~\cite{zeng2024colare} to classify commits into corrective, adaptive, and perfective categories by integrating hunk-level code representations with commit messages and file features. However, such models are build for commit-level classification which does not properly align with our method-level objectives.

In summary, this study is motivated by three factors: (1) the lack of fine-grained research on the tangled changes at the method level, especially for bug prediction~\cite{herzig_impact_2016, chowdhury_method-level_2024}; (2) the proven capabilities of LLMs in analyzing textual and code artifacts via advanced prompting~\cite{brown_language_2020, zheng2024few, wei_finetuned_2022, wei_chain--thought_2023}; and (3) the semantic richness of LLM-generated embeddings for classification~\cite{petukhova_text_2024, keraghel_beyond_2024}. \emph{To the best of our knowledge, this is the first study to evaluate LLMs for detecting tangled code changes within bug-fix commits at the method level.}
\section{Methodology}
\label{sec:methodology}

\begin{figure*}[ht]
    \centering
    \includegraphics[width=1\textwidth]{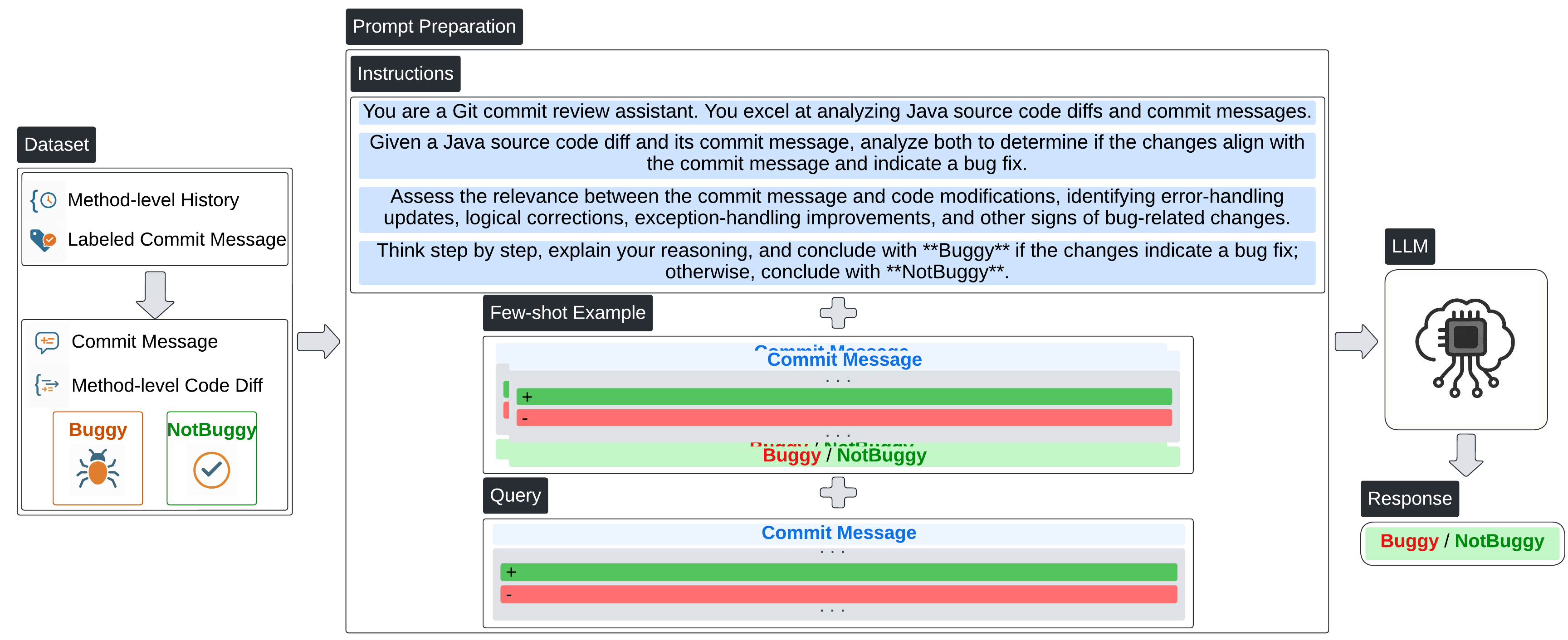}
    \caption{The methodology for \textit{RQ1} and \textit{RQ2} consists of three parts: (a) dataset, (b) prompt preparation, and (c) LLM Query. The prompt setup for few-shot + chain-of-thought prompting consists of four instruction parts: (i) persona, (ii) task description, (iii) behavioral guidance, and (iv) output formatting. For Diff-only detection of \textit{RQ1}, references to the commit message are excluded from everywhere. Other experiments use the full instructions, with the output formatting modified to require a single-word answer in non-chain-of-thought setups. Few-shot examples are included only in few-shot experiments.}
    \Description{Overview of methodology.}
    \label{fig:methodology}
\end{figure*}

Figure~\ref{fig:methodology} shows the overview of our approach that we use to evaluate LLMs in distinguishing \textit{Buggy} diff and \textit{NotBuggy} diff. In this section, the components of the methodology are described step by step.

%The dataset component also relates to \textit{RQ3}; the remaining methodology for \textit{RQ3} is discussed later.
\subsection{Dataset}
To assess whether LLMs can be used to untangle real bug fix changes from other changes, a dataset is required that includes both genuine bug fix code diffs and non-bug-fix code diffs. We provide the model with a code diff with or without the commit message to evaluate its ability to distinguish between bug-fix and non-bug-fix changes.

Fortunately, an existing dataset by Chowdhury et al.~\cite{chowdhury_method-level_2024} contains 774,051 Java methods from 49 open source projects, where each method is associated with its complete change history and is enriched with metadata such as commit messages, timestamps, authorship, and change type. The dataset also includes code diffs for all historical changes, labels indicating whether each change is related to a bug fix or not, and information on the number of methods modified within the same commit.

However, the dataset contains noise. In some cases, methods involved in a bug fix commit may not contribute to the actual fix, representing instances of tangled changes. We use this dataset to construct a curated gold dataset by removing such noise. The construction process involves a two-step sampling strategy, combining automated filtering and manual labeling, to ensure high quality and accurate representation of tangled change scenarios.

% To understand if LLM can be used to untangle real bugfix changes from other changes, we need a dataset where there are both real bugfix code diffs and non-real bugfix code diffs. Now we will provide LLM with this code diff and commit message to check if they can distinguish a bugfix change and a non-bugfix change. Fortunately, there is an existing dataset by Chowdhury et al.~\cite{chowdhury_method-level_2024}, comprising 774,051 Java methods across 49 open-source projects, where each method is linked to its full change history (as a sequence of diffs) and enriched with metadata, including commit messages, timestamps, authorship, and change type. 

% It also includes the code diff for all the changes throughout its history and which change is bug fix related and which one is not bug-fix related. Also, the number of methods changed in the same commit.

% This dataset is noisy as there may be some methods that changed in a bug fix commit but do not contribute to the bug fixed, i.e., a tangled change. We will use this dataset to make a curated gold dataset and later remove the noise from this dataset. The construction of our gold dataset followed a two-step sampling procedure, automated and manual,  designed to ensure both label quality and representation of tangled change scenarios.

\textbf{Automated Approach.} The dataset provided by Chowdhury et al.~\cite{chowdhury_method-level_2024} includes the total number of methods modified in each commit. They recommended that if a method is the only one modified in a bug fix commit, it can be confidently considered a bug-related change. We classify such code diffs as \textit{Buggy} without ambiguity. Using this criterion, we collected 730 Buggy diffs. They also suggested that a method can be considered \textit{NotBuggy} if it has never appeared in a bug fix commit throughout its lifetime. Based on this guideline and leveraging the large pool of eligible \textit{NotBuggy} methods, we randomly selected 730 samples that have no history of being part of any bug fix commit. During this process, three duplicate diffs, likely caused by duplicated code across codebases, were identified and removed, resulting in 727 unique NotBuggy instances. Thus, the \textit{Automated} step of our gold dataset construction produced a total of 1,457 method-level examples. We intentionally limited the number of NotBuggy samples, as querying proprietary LLMs incurs significant computational and financial costs. Including all such samples would have significantly increased the overall cost.

% \textbf{Automated Approach.}  The total number of methods modified in each commit is given in the dataset of Chowdhury et al., and in their work, they recommended that if a method is modified as the only method in a bug fix commit, then it is definitely a bug-related change. We can take this kind of code diff as \textit{Buggy} diff undoubtedly. in this approach we collected 730 Buggy diffs.
% They also recommended that a method should be treated as NotBuggy if it is not involved in any bug fix commits in its lifetime but they should be given enough time to evolve. Chowdhury et al. recommended the time as 2 years. So following this, given the large pool of eligible NotBuggy methods, we randomly collected 730 samples with age greater than 2 years and never associated with any bugfix commit as NotBuggy samples. Among them, 3 duplicate diffs, likely due to duplicated code in the codebases, were identified and removed, yielding 727 unique instances. As a result, \textit{Phase 1} of our dataset construction yielded a total of 1,457 method-level examples. At the same time we did not take all the NotBuggy samples as using LLM is expensive. Taking all the NotBuggy samples will increase the cost substantially.

\textbf{Manual Approach.} While the \textit{Automated} approach ensured high confidence in labeling, it did not capture the complexity of real-world tangled change scenarios, where multiple unrelated modifications often occur within a single commit. To address this limitation, we introduced a second step focused on incorporating tangled examples. In this step, we extracted additional method-level diffs from bug-fix commits that modified multiple methods. The first author, a graduate student with over two years of industry experience in software engineering, manually labeled each method-level change to determine whether it was related to a bug fix (\textit{Buggy}) or was incidental, such as refactorings or unrelated enhancements (\textit{NotBuggy}) Each method was reviewed in conjunction with its associated commit message to evaluate semantic alignment with bug-fixing intent. The evaluation emphasized structural corrections, error handling logic, and domain-specific repairs. To promote annotation consistency and minimize errors, we developed a user interface that allowed side-by-side inspection of code diffs and their corresponding commit metadata. Using this \textit{manual} approach, we identified 166 tangled \textit{Buggy} methods and 141 tangled \textit{NotBuggy} methods from 450 randomly selected commits in the dataset of Chowdhury et al.~\cite{chowdhury_method-level_2024}.

% \textbf{Manual Approach.} While \textit{Automated} approach ensured high-confidence labels, it did not reflect the complexity of real-world tangled change scenarios, where multiple unrelated modifications often co-occur in a single commit. To capture this nuance, we initiated a second step focused on incorporating tangled examples. Here, we extracted additional method-level diffs from multi-method commits and conducted manual annotations to determine whether each method-level change was genuinely related to a bug fix (Buggy) or was incidental, such as refactorings or unrelated enhancements (NotBuggy). Each method was reviewed alongside its commit message to assess semantic alignment with bug-fixing intent, emphasizing structural corrections, error-handling logic, and domain-relevant repairs. To support consistency and reduce annotation error, we developed a dedicated user interface that enabled side-by-side inspection of code diffs and their corresponding commit metadata. In \textit{Manual} approach, a total of 166 Tangled Buggy methods and 141 Tangled NotBuggy methods were collected through manual annotation. 

Combining all steps, the gold dataset consists of 1,764 method-level change instances, each labeled as either \textit{Buggy} or \textit{NotBuggy}, based on their actual relevance to bug fixing activities. To evaluate labeling reliability, an independent validation was conducted by another graduate student in software engineering, who randomly picked 100 samples from the manually labeled portion of the dataset and labeled them independently. Some disagreements arose, mainly due to differing interpretations of what constitutes a bug. Specifically, the first labeler marked some quality-related fixes as \textit{NotBuggy}, while the second considered them \textit{Buggy}. Following a discussion with the last author, quality-fix-related changes were finalized as \textit{NotBuggy}. The inter-rater agreement, measured using Cohen’s Kappa~\cite{cohen1960coefficient}, was 0.82, indicating strong consistency and confirming the reliability of the labeling process. Consequently, this dataset provides a robust benchmark for both training and evaluating our detection approaches, while also offering a valuable foundation for future research in this domain.

\subsection{Prompt Preparation}
The design of our prompt structure draws on recent advancements in prompt engineering that demonstrate the effectiveness of decomposing complex instructions into clearly scoped components~\cite{brown_language_2020, wei_chain--thought_2023, zhou_least--most_2023}. As prior work has shown, LLMs are highly sensitive to prompt formulation, and performance can vary substantially based on how the task, instructions, and examples are framed~\cite{liu_pre-train_2023}. To maximize reliability and adaptability, we designed a modular prompt template consisting of three core components, aligned with prior studies~\cite{amatriain_prompt_2024, giray_prompt_2023}, as illustrated in Figure~\ref{fig:methodology}.

\begin{figure*}[ht]
    \centering
    \includegraphics[width=0.98\linewidth]{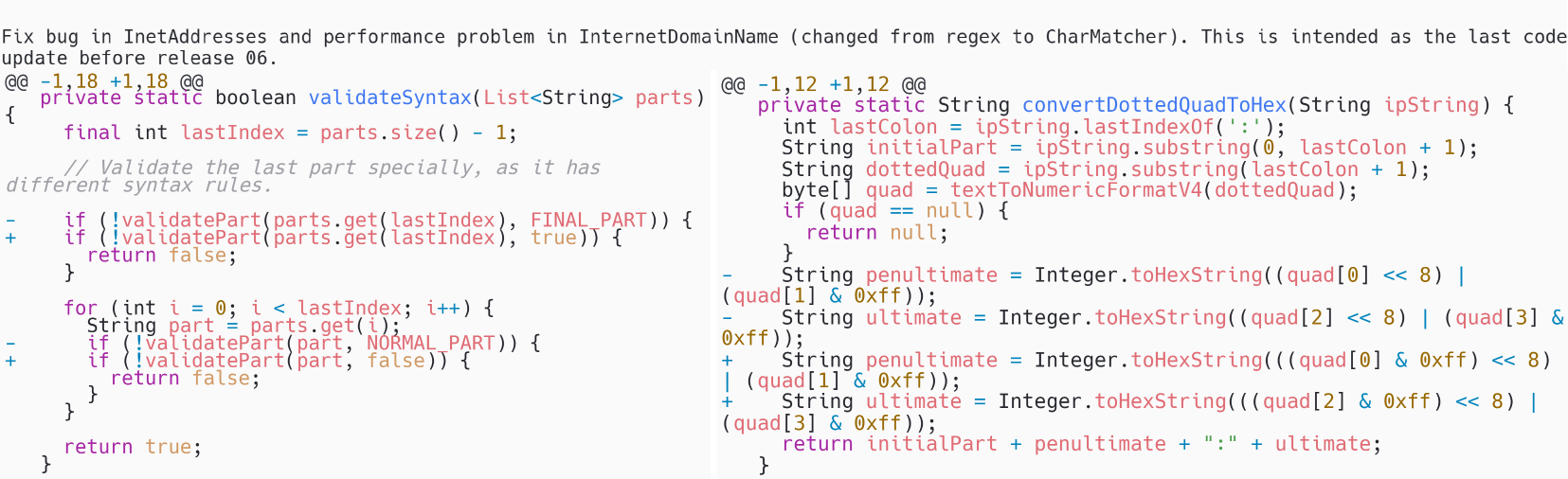}
    \caption{Commit message and two methods modified in the same commit: the message indicates a bug fix, but only the method on the right contains the bug-fix change, while the method on the left is a non-bug-related tangled change. This was used as few-shot examples in our prompts.}
    \Description{diff}
    \label{fig:diff}
\end{figure*}

\textbf{Instructions.} This component defines the LLM's role and behavior. It is composed of four parts: (i) a role-based persona (e.g., “You are a Git commit review assistant”), which helps constrain the model's responses by grounding its behavior and improves performance~\cite{hu_quantifying_2024}; (ii) a high-level task description~\cite{amatriain_prompt_2024}; (iii) step-by-step behavioral guidance to structure the model's reasoning~\cite{wei_chain--thought_2023}; and (iv) explicit output formatting requirements to reduce ambiguity and improve parsability~\cite{liu_are_2024}.

\textbf{Few-shot Examples.} Few-shot prompting has been shown to significantly improve performance by enabling in-context learning, especially in code and software engineering tasks~\cite{zheng2024few}. For few-shot examples, as shown in Figure~\ref{fig:diff}, we selected a representative bug-fix commit containing one diff genuinely related to a bug fix and another unrelated change, a real instance of tangled code changes. For the negative example (\textit{NotBuggy}), we selected the code diff that merely replaced constant values with equivalent boolean literals, without altering the program’s logic or behavior. In contrast, the other code diff (\textit{Buggy}) addressed an issue in the byte-to-hexadecimal conversion logic by correctly handling signed bytes using bit masking, rectifying a functional error. These two fixed examples were consistently utilized to prime the LLMs in all experiments employing few-shot learning. These examples were selected to represent contrasting scenarios, one clearly indicative of a bug-fix and the other representing a refactor or performance-related change that does not qualify as a bug-fix, although both of the examples come from the same commit.

LLMs often exhibit significant performance degradation as prompt length increases, due to limitations in handling extended context~\cite{levy_same_2024, zhang_inftybench_2024}. Additionally, longer prompts incur higher computational and monetary costs, making token efficiency a practical concern. Since method-level diffs can vary greatly in length and the input query, over which we have no control, is the primary driver of prompt size, we deliberately restricted the prompt to a single example per class (\texttt{Buggy} and \texttt{NotBuggy}). This prompting approach helps maintain prompt lengths within manageable token limits across different model families. By limiting the number of in-context examples, we strike a balance between performance and resource usage. This strategy aligns with established practices in few-shot prompting ~\cite{brown_language_2020}, where concise and diverse instances are used to anchor the model’s reasoning prior to inference. 
%In our setup, the examples provide contextual grounding by showcasing typical linguistic and structural patterns present in both buggy and non-buggy commits.

\textbf{Query.} This includes the method-level code diff and its corresponding commit message.

Each research question required slight variations in prompt configuration, e.g., enabling or disabling few-shot examples, toggling chain-of-thought reasoning, or isolating input modalities (diff-only vs. diff + message). All corresponding prompt templates are included in our replication package, located within the \textit{prompts} directory.

\subsection{Selecting and Querying LLMs}

\textbf{Selecting LLMs.} For RQ1 and RQ2, we selected six LLMs: four proprietary, \textit{GPT-4o-mini}, \textit{GPT-4o}, \textit{GPT-5}, and \textit{Gemini-2.0-Flash} and two open-source, \textit{GPT-OSS-20B} and \textit{GPT-OSS-120B}. The proprietary LLMs were chosen based on their performance on recent code-related benchmarks~\cite{google2024geminiupdate, openai2025gpt4omini, batista_code_2024, bruni_benchmarking_2025}, robust API support, and cost-effectiveness for large-scale experimentation. \textit{GPT-4o} serves as OpenAI’s full-scale model offering superior robustness over its smaller variant, \textit{GPT-4o-mini}~\cite{openai2025gpt4omini, bruni_benchmarking_2025}, while \textit{GPT-5} represents the latest generation of OpenAI’s LLMs. \textit{Gemini-2.0-Flash}, from Google~\cite{google2024geminiupdate}, delivers competitive performance with the GPT series and has demonstrated effectiveness in code-related tasks~\cite{batista_code_2024}.

For open-source alternatives and to ensure reproducibility and accessibility, we choose \textit{GPT-OSS-20B} (requiring ~16 GB GPU memory) and \textit{GPT-OSS-120B} (runnable on a single 80 GB GPU)~\cite{agarwal2025gpt}. Despite their moderate scale, these models deliver reasoning and code-understanding performance comparable to proprietary baselines~\cite{agarwal2025gpt, bi2025gpt}, balancing efficiency and capability.

For RQ3, we employed four models for embedding generation: one proprietary and three open-source. The proprietary model, Gemini's \textit{text-embedding-004} offers high-quality embeddings optimized for classification tasks~\cite{lee2024gecko}. Open-source models, \textit{CodeBERT}~\cite{feng2020codebert} and \textit{GraphCodeBERT}~\cite{guo2020graphcodebert} are widely adopted for code representation learning~\cite{ma2024unveiling}, while \textit{EmbeddingGemma-300m}, a recent lightweight model from Google DeepMind, is designed for low-latency, high-throughput applications~\cite{vera2025embeddinggemma}. The models used in RQ1 and RQ2 were not included in RQ3 because they are decoder-only generative models optimized for text generation, whereas RQ3 required encoder-based architectures specialized for producing fixed-length embeddings suitable for classification tasks. 

In summary, proprietary models were selected for their state-of-the-art performance across benchmarks, including code-related tasks~\cite{batista_code_2024, bae2024enhancing}, while open-source models were selected for their competitive accuracy and efficiency relative to larger open-source alternatives. 

% If our selected LLMs cannot solve the tangling problem, it is unlikely that other models can. Their performance is likely to be the upper bound for all LLMs. However, we acknowledge that more experiments with other LLMs can produce interesting results.

\textbf{Querying LLMs.} The three parts of the prompt (instruction + examples + query) are concatenated and formatted according to the input constraints of the respective LLM. The final query prompt is submitted to the LLM, which classifies the change as either \textit{Buggy} or \textit{NotBuggy}. Outputs are reviewed and filtered automatically, if required, to ensure alignment with evaluation criteria.

\subsection{Evaluation Metrics}
We evaluate model performance using standard classification metrics, including accuracy, precision, recall, and F1-score. Additionally, we use the Matthews Correlation Coefficient (MCC), a robust metric that considers all four outcomes: true positives, true negatives, false positives, and false negatives. Unlike accuracy, MCC provides a more balanced measure. An MCC of 1 indicates perfect prediction, 0 indicates random guessing, and –1 reflects total disagreement between predictions and actual labels.

\subsection{Experimental Setup}
All experiments involving open-source LLMs were conducted on a High-Performance Computing (HPC) cluster equipped with Intel 8570 CPUs (2.1 GHz) and multiple NVIDIA H100 GPUs. The number of CPUs, GPUs and the memory allocation varied by model, depending on resource availability and scheduling within the cluster.

% \begin{itemize}
%     \item \textbf{Precision:} Measures the proportion of correctly identified \texttt{Buggy} instances among all instances classified as \texttt{Buggy}.
%     \item \textbf{Recall:} Measures the proportion of actual \texttt{Buggy} instances that were correctly identified.
%     \item \textbf{F1-Score:} Harmonic mean of precision and recall, providing a balanced metric.
%     \item \textbf{Accuracy:} Overall proportion of correctly classified instances.
%     \item \textbf{Matthews Correlation Coefficient (MCC):} A robust metric accounting for true and false positives and negatives, especially useful in imbalanced datasets. It also shows whether the predictions are just random. MCC is 0 for random prediction and 1 for perfect prediction. 
% \end{itemize}
%\input{Sections/4_DataCollection}
\section{Approach, Analysis \& Results}
\label{sec:results}
This section answers the four research questions and outlines the specific methodologies used to investigate each one.

\subsection{RQ1: Performance of zero-shot LLMs using code diff and commit message}

Both the commit message and the corresponding code diff contain information about the nature of the code changes. While the diff captures the structural modifications to the source code, the commit message provides a natural language description of the developer’s intent. LLMs offer a powerful mechanism to detect semantic signals indicative of bug-fixing behavior across these modalities. By leveraging their ability to reason over structured and unstructured data (code + natural language)~\cite{nam_using_2024, chew_llm-assisted_2023}, LLMs offer a promising alternative to traditional rule-based or static analysis techniques. Furthermore, using prompt-based inference avoids the need for task-specific fine-tuning, making the approach adaptable across repositories and languages. However, the effectiveness of LLMs is sensitive to input length, as inference cost increases with the number of tokens in the prompt. Given that both code diffs and commit messages can vary significantly in length, it becomes essential to evaluate whether both are necessary for effective detection or whether the code diff alone suffices. This research question seeks to empirically determine whether commit messages provide additive value beyond the code diff for detecting tangled changes.

We investigate two scenarios using LLMs by designing tailored prompting strategies for the binary classification of method-level diffs. Since no examples are provided within the prompt, this setup qualifies as zero-shot prompting.

% \begin{itemize}
%   \item 
\textbf{Diff-Only.} The LLM is provided only with the code diff and instructed to assess whether the change represents a bug fix (\texttt{Buggy}) or not (\texttt{NotBuggy}). The prompt is adapted from Figure~\ref{fig:methodology}, with all references to commit messages and few-shot examples removed, and the instruction modified to require a single-word response.
% A representative fragment of the prompt is: "\textit{... Given a Java source code diff, analyze it to determine if the changes are related to a bug fix. Examine the modifications carefully, looking for error-handling ...}". The output formatting is also modified to: "\textit{Output only a single word: 'Buggy' if the changes indicate a bug fix, otherwise 'NotBuggy'.}", since the format in Figure~\ref{fig:methodology} is intended for chain-of-thought.

\textbf{Diff+Message.} The prompt includes both the code diff and its commit message, instructing the model to assess their semantic alignment to identify bug-fixing intent. This prompt follows the \textit{Diff-Only Detection} setting, except that all references to commit messages are retained as in Figure~\ref{fig:methodology}.
%\end{itemize}

% Table~\ref{tab:rq1-llm-performance} compares LLM performance under two input scenarios: Diff-only and Diff+Message. In the Diff-only setting, \texttt{gpt-4o-mini} achieves the highest performance across all metrics, with an F1 score of 0.747. When commit messages are included, all models show substantial improvements. \texttt{gpt-4o} yields the best overall results, attaining the highest accuracy (0.874), F1 score (0.877), and MCC (0.747). Although \texttt{gemini-2.0-flash} achieves the highest recall (0.988), its low precision (0.717) leads to a reduced F1 score (0.831). These findings indicate that incorporating commit messages alongside diffs consistently enhances defect detection performance.

Table~\ref{tab:rq1-llm-performance} compares the performance of the evaluated LLMs under two input settings: \textit{Diff-only} and \textit{Diff+Message}. In the \textit{Diff-only} scenario, \textit{gpt-5} achieves the best overall performance, yielding the highest F1-score (0.767) and MCC (0.490). When commit messages are included, all models show substantial improvements. The \textit{gpt-5} again attains the highest F1-score (0.879), while \textit{gpt-4o} provides the best accuracy (0.874), precision (0.868) and MCC (0.747). Although \textit{gpt-5} achieves the top F1-score, its high precision–recall gap indicates less balanced performance, whereas \textit{gpt-4o} offers the most consistent results across all metrics.

The \textit{gemini-2.0-flash} and the \textit{gpt-oss} models achieve high recall ($>0.9$) but low precision ($<0.76$) resulting in reduced F1-scores. The \textit{gpt-oss} models perform remarkably well in \textit{Diff+Message} scenario despite their smaller size narrowing the gap with proprietary models (within ~3–4\% of \textit{gpt-5}’s F1-score), likely benefiting from their unique dual-channel chat format~\cite{agarwal2025gpt}, where an internal “analysis” channel generates intermediate reasoning (chain-of-thought) before producing the answer on "final" channel. Overall, these results demonstrate that integrating commit messages with diffs significantly improves performance, and that open-source LLMs are also effective despite their smaller sizes.

\begin{table}[h]
\caption{Performance comparison of LLMs for Diff-only and Diff+Message. Metrics are calculated by targeting the buggy samples (pos\_label = "Buggy"). For each scenario, the best value for each metric is highlighted in bold.}
\label{tab:rq1-llm-performance}
\resizebox{\columnwidth}{!}{%
\addtolength{\tabcolsep}{-0.25em}
\begin{tabular}{@{}ccccccc@{}}
\toprule
\rotatebox{0}{\textbf{Scenario}} & 
\rotatebox{0}{\textbf{LLM}} & 
\rotatebox{0}{\textbf{Accuracy}} & 
\rotatebox{0}{\textbf{Precision}} & 
\rotatebox{0}{\textbf{Recall}} & 
\rotatebox{0}{\textbf{F1-score}} & 
\rotatebox{0}{\textbf{MCC}} \\
\midrule
\multirow{6}{*}{Diff-only}
 & gpt-5 & \textbf{0.742} & \textbf{0.709} & 0.834 & \textbf{0.767} & \textbf{0.490} \\
 & gpt-4o-mini & 0.704 & 0.660 & 0.860 & 0.747 & 0.426 \\
 & gpt-4o & 0.691 & 0.654 & 0.831 & 0.732 & 0.395 \\
 & gemini-2.0-flash & 0.692 & 0.650 & 0.853& 0.738 & 0.402 \\
 & gpt-oss-20b & 0.663 & 0.615 & \textbf{0.900} & 0.730 & 0.364 \\
 & gpt-oss-120b & 0.667 & 0.619 & \textbf{0.900} & 0.733 & 0.372 \\
\midrule
\multirow{6}{*}{\makecell{Diff\\+\\Message}} 
 & gpt-5 & 0.871 & 0.839 & 0.924 & \textbf{0.879} & 0.746 \\
 & gpt-4o-mini & 0.822 & 0.764 & 0.941 & 0.843 & 0.661 \\
 & gpt-4o & \textbf{0.874} & \textbf{0.868} & 0.886 & 0.877 & \textbf{0.747} \\
 & gemini-2.0-flash & 0.796 & 0.717 & \textbf{0.988} & 0.831 & 0.639 \\
 & gpt-oss-20b & 0.820 & 0.758 & 0.950 & 0.843 & 0.662 \\
 & gpt-oss-120b & 0.827 & 0.766 & 0.948 & 0.847 & 0.671 \\
\bottomrule
\end{tabular}%
}
\end{table}

\begin{summarybox}
\textbf{Summary of RQ1}: Both proprietary and open-source LLMs can effectively detect tangled code changes using code diffs, with or without commit messages, though combining both inputs yields better results than code diffs alone.
\end{summarybox}

\subsection{RQ2: Performance of different prompting techniques}

The effectiveness of LLMs highly depends on the input prompt, and the success of different prompting techniques is proven in various studies~\cite{wei_finetuned_2022, brown_language_2020, wei_chain--thought_2023}. This research question investigates how different prompting techniques, namely \textit{few-shot learning}, \textit{chain-of-thought prompting}, and a hybrid approach combining both, impact the classification performance of LLMs.

% We implemented the following prompting techniques:

% \begin{itemize}
%   \item 
\textbf{Few-shot.} The prompt includes labeled examples of buggy and non-buggy changes to guide the model, while the instruction remains identical to the \textit{Diff+Message} setting of RQ1.

\textbf{Chain-of-thought.} The model is instructed to explain its reasoning in the output. The exact instruction prompt is shown in Figure~\ref{fig:methodology}, and no examples are included.

\textbf{Chain-of-thought + Few-shot.} This setting combines the above two techniques by including labeled examples and requiring the model to explain its reasoning in the output.
%\end{itemize}

Based on the findings from RQ1, we use the combination of commit message and code diff for RQ2. Additionally,  we exclude \textit{gpt-4o-mini} from further experiments, as \textit{gpt-4o} significantly outperformed this small model. This decision also supports cost efficiency, as the prompting techniques used in RQ2 involve substantial token consumption in both input and output. 

Table~\ref{tab:rq2-llm-performance} presents the performance of different prompting strategies across LLMs. While \texttt{gpt-4o} with few-shot achieves the highest precision (0.903) but lower recall (0.833), other models exhibit the opposite trend. In this setting, \textit{gpt-5} attains the best overall performance with an F1-score of 0.884. For chain-of-thought prompting, performance generally decreases across models, though \textit{gpt-5} again achieves the highest F1-score (0.872). When combining few-shot and chain-of-thought prompting, \textit{gpt-4o} delivers the best results across all metrics except recall, which is highest for \textit{gemini-2.0-flash}. 

\begin{table}[h]
\caption{Performance comparison of different prompting techniques. Metrics are calculated by targeting the buggy samples (pos\_label = "Buggy").}
\label{tab:rq2-llm-performance}
\centering
\resizebox{\columnwidth}{!}{%
\addtolength{\tabcolsep}{-0.3em}
\begin{tabular}{@{}ccccccc@{}}
\toprule
\rotatebox{0}{\makecell{\textbf{Prompting}\\\textbf{Technique}}} & 
\rotatebox{0}{\textbf{LLM}} & 
\rotatebox{0}{\textbf{Accuracy}} & 
\rotatebox{0}{\textbf{Precision}} & 
\rotatebox{0}{\textbf{Recall}} & 
\rotatebox{0}{\textbf{F1-score}} & 
\rotatebox{0}{\textbf{MCC}} \\
\midrule
\multirow{5}{*}{Few-shot} 
& gpt-5 & \textbf{0.876} & 0.842 & 0.931 & \textbf{0.884} & \textbf{0.757} \\
& gpt-4o & 0.870 & \textbf{0.903} & 0.833 & 0.866 & 0.742 \\
& gemini-2.0-flash & 0.802 & 0.723 & \textbf{0.989} & 0.835 & 0.649 \\
& gpt-oss-20b & 0.837 & 0.775 & 0.958 & 0.857 & 0.694 \\
& gpt-oss-120b & 0.819 & 0.753 & 0.959 & 0.843 & 0.663 \\
\midrule
\multirow{5}{*}{\makecell{Chain\\of\\thought}} 
& gpt-5 & \textbf{0.863} & \textbf{0.829} & 0.920 & \textbf{0.872} & \textbf{0.730} \\
& gpt-4o & 0.827 & 0.770 & 0.940 & 0.846 & 0.669 \\
& gemini-2.0-flash & 0.765 & 0.688 & 0.984 & 0.810 & 0.587 \\
& gpt-oss-20b & 0.829 & 0.770 & \textbf{0.946} & 0.849 & 0.676 \\
& gpt-oss-120b & 0.829 & 0.765 & 0.956 & 0.850 & 0.679 \\
\midrule
\multirow{5}{*}{\makecell{\textbf{Chain}\\\textbf{of}\\\textbf{thought}\\\textbf{+}\\\textbf{Few-shot}}}
& gpt-5 & 0.863 & 0.829 & 0.921 & 0.873 & 0.731 \\
& \textbf{gpt-4o} & \textbf{0.880} & \textbf{0.871} & 0.896 & \textbf{0.883} & \textbf{0.760} \\
& gemini-2.0-flash & 0.807 & 0.731 & \textbf{0.979} & 0.837 & 0.651 \\
& gpt-oss-20b & 0.844 & 0.795 & 0.932 & 0.858 & 0.697 \\
& gpt-oss-120b & 0.816 & 0.752 & 0.952 & 0.840 & 0.655 \\
\bottomrule
\end{tabular}%
}
\end{table}

Among all prompting strategies, \textit{gpt-5} with few-shot yields the top F1-score, while \textit{gpt-4o} attains the highest accuracy, precision, and MCC with chain-of-thought+few-shot. Considering balanced performance across metrics, particularly the small gap between precision and recall (0.025), \textit{gpt-4o} with chain-of-thought+few-shot offers the most balanced results, with an F1-score (0.883) nearly identical to the best (0.884). Open-source models also benefit consistently, though modestly, from advanced prompting; notably, \textit{gpt-oss-20b} remains competitive, often matching or surpassing \textit{gpt-oss-120b}.

A comparison between Table~\ref{tab:rq1-llm-performance} and Table~\ref{tab:rq2-llm-performance} reveals that few-shot prompting improves performance primarily for \textit{gpt-5} and slightly for \textit{gpt-oss-20b}, while other models show marginal or decreased performance. Notably, chain-of-thought alone results in poorer performance across most models. However, combining few-shot prompting with chain-of-thought yields the most balanced and consistent outcomes overall, as observed with \textit{gpt-4o}. This trend aligns with prior research suggesting that few-shot prompting can lead models to overfit to example formats rather than the task itself, while zero-shot chain-of-thought prompting may amplify biases and degrade reasoning quality~\cite{kojima_large_nodate, shaikh_second_2023}. Our findings thus corroborate earlier studies indicating that integrating few-shot and chain-of-thought generally outperforms zero-shot approaches~\cite{kojima_large_nodate}.

%For example, when few-shot examples vary in answer types, the resulting performance often degrades below that of zero-shot prompting. Although chain-of-thought can improve reasoning in some situations, zero-shot chain-of-thought has also been shown to produce unintended side effects.

\renewcommand{\arraystretch}{1.3}
\begin{table*}[ht]
\small
\caption{Predictions on a semantically misleading commit using chain-of-thought + few-shot prompting with \textit{gpt-4o}.}
\label{tab:misclassification-case}
\centering
\begin{tabular}{|p{4cm}|p{5cm}|c|c|c|}
\hline
\textbf{Commit Message} & \textbf{Diff (Excerpt)} & \makecell{\textbf{Ground}\\\textbf{Truth}} & \makecell{\textbf{Prompting}\\\textbf{Method}} & \textbf{Prediction} \\
\hline
\multirow{4}{=}{Fixed bug 1050173 – \textit{ImmutableFieldRule} no longer reports false positives for static fields. Also fixed version number in PMD.java.} & 
\multirow{4}{=}{%
\begin{minipage}[t]{5cm}
\small\textit{
...\\
- NameOccurrence occurance = ...\\
- if (occurance.isOnLeftHandSide()) \{\\
\ \ \ \ ...\\
+ NameOccurrence occ = ...\\
+ if (occ.isOnLeftHandSide()) \{ ...\\
}
\end{minipage}
}
& \multirow{4}{*}{NotBuggy} & \makecell{Commit msg + diff\\(zero-shot)} & Buggy \\
\cline{4-5}
& & & Few-shot & Buggy \\
\cline{4-5}
& & & Chain-of-thought & \textbf{NotBuggy} \\
\cline{4-5}
& & & \makecell{Chain-of-thought +\\Few-shot} & \textbf{NotBuggy} \\
\hline
\end{tabular}
\end{table*}
\renewcommand{\arraystretch}{1} 

\subsubsection{\textbf{Case analysis: an example of the effectiveness of chain-of-thought approach in a semantically ambiguous commit}}
To better understand why the chain of thought technique improves model performance in the few shot setting with \textit{gpt-4o}, we analyze a representative example involving semantic ambiguity between the commit message and the code modifications from the \textit{PMD} \footnote{\url{https://github.com/pmd/pmd}} project. As shown in Table~\ref{tab:misclassification-case}, the commit message clearly describes a bug fix involving two files, \textit{PMD.java} and \textit{ImmutableFieldRule.java}. In this commit\footnote{\url{https://github.com/pmd/pmd/commit/a405d23dfb9e574e2b2ef23f1f45d548a738ed3b}}, multiple methods are modified, including \textit{initializedInConstructor()} from \textit{ImmutableFieldRule.java}, which only includes variable renaming (e.g., \textit{occurance} to \textit{occ}) without affecting the program’s logic or behavior. No structural fix is present that addresses the reported issue. Despite this, both the zero-shot and few-shot models incorrectly classify the change as \textit{Buggy}, influenced by the commit message. In contrast, the chain-of-thought technique, with or without few-shot prompting, correctly identifies the change as \textit{NotBuggy}, consistent with the manually assigned label.

The reasoning produced by the chain-of-thought prompt illustrates a more grounded assessment. An important part of the reasoning states: {\itshape
\noindent ". . . However, in the provided diff, the only change is renaming the variable \textit{occurance} to \textit{occ}, which is a simple refactor for readability or consistency purposes. This modification itself doesn't address the logic or behavior of the code in terms of resolving the stated bug. The commit message also mentions a version number fix in \textit{PMD.java}, but no relevant changes are shown in this diff. . . ."
}

This case highlights the effectiveness of chain-of-thought in systematically analyzing complex commit messages that reference multiple issues. While such messages offer valuable context, they can also introduce ambiguity. The step-by-step reasoning decomposed the message, independently assessed each part against the code diff, and correctly identified the lack of structural or semantic alignment, ultimately concluding that the change was not a bug fix.

\begin{summarybox}
\textbf{Summary of RQ2}: Combining chain-of-thought reasoning with few-shot prompting yields the most balanced and consistent performance across models. While few-shot prompting alone benefits mainly \textit{gpt-5}, and chain-of-thought alone often reduces accuracy, their combination enhances reasoning stability and mitigates overfitting. This hybrid approach also improves interpretability, enabling models to correctly handle semantically misleading commits through more context-aware reasoning.
\end{summarybox}

\subsection{RQ3: Performance of embedding-based machine learning models}

This research question explores the viability of embedding-based ML models for detecting tangled code changes. Unlike prompt-based approaches that rely on in-context reasoning of LLMs, this method uses embeddings, vector representations of the input data from commit messages and code diffs, to train supervised classifiers. The key objective is to assess whether these embeddings, when used as features in ML models, can achieve competitive performance in detecting bug-related code changes. This experiment is motivated by the previous studies where embedding-based models performed well and even outperformed prompt-based techniques~\cite{petukhova_text_2024, keraghel_beyond_2024}. 

\begin{figure*}[ht]
    \centering
    \includegraphics[width=0.88\textwidth]{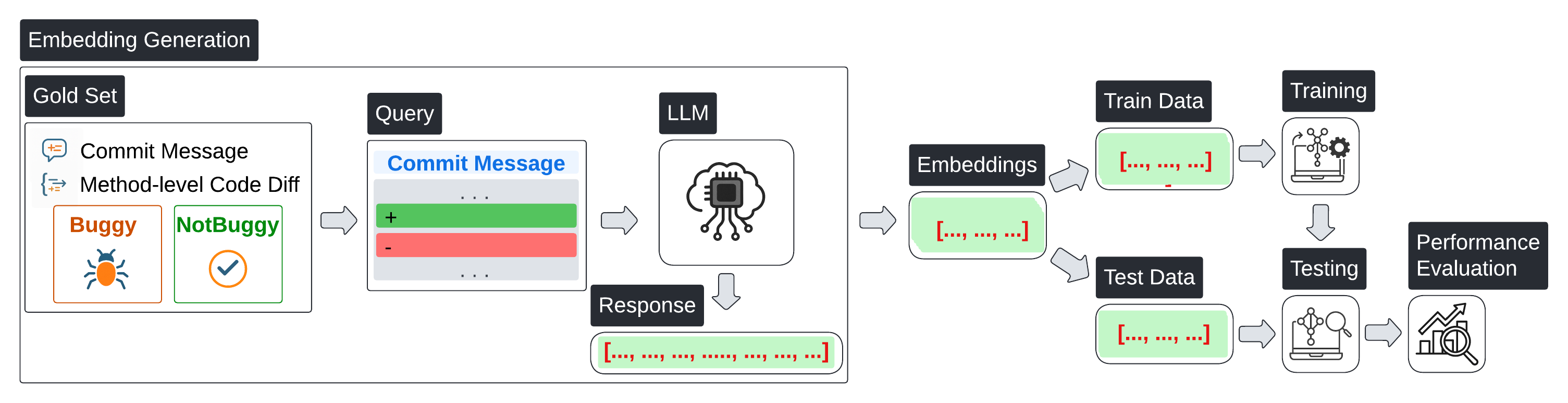}
    \caption{The methodology for RQ3 has two main parts: (i) embedding generations and (ii) building ML models.}
    \Description{Overview of methodology for embedding generation.}
    \label{fig:embedding_methodology}
\end{figure*}

The methodology employed for this research question diverges from that of the preceding research questions and the process illustrated in Figure~\ref{fig:methodology}. Specifically, this experiment omits the complex prompt engineering. Instead, only the core query, comprising the code diff and commit message, is retained and directly input to the LLM to generate fixed-length vector embeddings for downstream classification. The modified methodology is shown in Figure~\ref{fig:embedding_methodology}.

To generate embeddings, Gemini’s \textit{text-embedding-004} was employed with the \textit{classification} task type. Due to resource constraints, this study focuses exclusively on this proprietary model, utilizing Gemini’s free quota. For the open-source models, a sliding-window approach with mean pooling was applied to address token input limits, 2048 for \textit{EmbeddingGemma-300m} and 512 for \textit{CodeBERT} and \textit{GraphCodeBERT}. All models produced 768-dimensional vector representations for each instance, derived from the concatenation of the code diff and its corresponding commit message.

% \textcolor{blue}{To generate embeddings, we employed \texttt{text-embedding-004} with the \textit{Classification} task type, \texttt{EmbeddingGemma-300m}, \texttt{CodeBERT}, and \texttt{GraphCodeBERT}. For \texttt{CodeBERT} and \texttt{GraphCodeBERT}, we applied a sliding-window approach with mean pooling to address the 512-token input limitation. These models yields 768-dimensional vector representations for each input instance, constructed by concatenating the code diff and its commit message. Due to constraints in resources, this study focuses exclusively on this proprietary LLM. To reduce cost, other proprietary models were excluded, and Gemini's free quota was used to generate the embeddings. We acknowledge that we may achieve higher accuracy using other models. Nonetheless, as the objective is to assess the general effectiveness of embedding-based models, the results offer a representative indication of the potential performance of similar LLM embeddings.}

The classification task was treated as a binary problem, and the dataset was split into an 80\% training set and a 20\% test set. We evaluated multiple classifiers, including RandomForest (RF), Support Vector Classifier (SVC), DecisionTree (DT), and Multi-Layer Perceptron (MLP). These models were selected to represent a diverse range of learning paradigms: tree-based ensembles (RF, DT), margin-based optimization (SVC), and neural networks (MLP), which have been widely applied in various software engineering tasks~\cite{pascarella_performance_2020, chowdhury_method-level_2024}. 

\begin{table}[H]
\caption{Performance comparison of ML models trained on \textit{text-embedding-004} embeddings. Metrics are calculated by targeting the buggy samples (pos\_label = "Buggy").}
\label{tab:rq3-gemini-classifier-comparison}
\centering
\resizebox{\columnwidth}{!}{%
\addtolength{\tabcolsep}{-0.2em}
\begin{tabular}{lccccc}
\toprule
\textbf{Classifier} & \textbf{Accuracy} & \textbf{Precision} & \textbf{Recall} & \textbf{F1-score} & \textbf{MCC} \\
\midrule
RF & 0.816 & 0.841 & 0.82 & 0.83 & 0.629 \\
SVC & 0.884 & 0.918 & 0.866 & 0.891 & 0.768 \\
DT & 0.686 & 0.717 & 0.706 & 0.712 & 0.366 \\
\textbf{MLP} & \textbf{0.895} & \textbf{0.929} & \textbf{0.876} & \textbf{0.902} & \textbf{0.791} \\
\bottomrule
\end{tabular}
}
\end{table}

Table~\ref{tab:rq3-gemini-classifier-comparison} presents the performance comparison of all models trained on \textit{text-embedding-004} embeddings, showing that the \textit{MLPClassifier} achieved the best results, with an accuracy of 0.895, F1-score of 0.902, and MCC of 0.791, outperforming other classifiers in all metrics. Similar experiments were conducted using open-source embeddings; however, due to space limitations, only the MLP results are reported in Table~\ref{tab:rq3-classifier-comparison}, as it consistently outperformed other classifiers. Among the open-source models, \textit{GraphCodeBERT} combined with MLP achieved the highest performance (F1-score = 0.831), exhibiting a modest gap of around 7\% compared to the proprietary model. This suggests that while proprietary embeddings retain a advantage, open-source alternatives offer competitive and cost-effective performance for this task.

\begin{table}[H]
\caption{Performance comparison of MLP classifier trained on different LLM embeddings. The first row of  \textit{text-embedding-004} is taken from Table~\ref{tab:rq3-gemini-classifier-comparison} for comparison. Metrics are calculated by targeting the buggy samples (pos\_label = "Buggy").}
\label{tab:rq3-classifier-comparison}
\centering
\resizebox{\columnwidth}{!}{%
\addtolength{\tabcolsep}{-0.3em}
\begin{tabular}{lccccc}
\toprule
\textbf{Model} & \textbf{Accuracy} & \textbf{Precision} & \textbf{Recall} & \textbf{F1-score} & \textbf{MCC} \\
\midrule
text-embedding-004 & \textbf{0.895} & \textbf{0.929} & \textbf{0.876} & \textbf{0.902} & \textbf{0.791} \\
EmbeddingGemma & 0.80 & 0.815 & 0.82 & 0.818 & 0.593 \\
CodeBERT & 0.765 & 0.738 & 0.833 & 0.782 & 0.533 \\
GraphCodeBERT & 0.828 & 0.831 & 0.83 & 0.831 & 0.656 \\
\bottomrule
\end{tabular}
}
\end{table}

Furthermore, we evaluated the \texttt{MLPClassifier} using the \textit{text-embedding-004} embeddings under a Leave-One-Out (LOO) strategy to test its robustness, where a single example is used as the test sample and the remaining serve as the training data. This approach not only offers more robust evaluation, but also can show if a model's accuracy can be improved with more training samples. Encouragingly, the model achieved improved performance with an accuracy of 0.9036, precision of 0.8980, recall of 0.9141, F1-score of 0.906, and MCC of 0.8073, setting a new benchmark compared to prior best results in RQ1 and RQ2. In terms of class-wise performance the model maintained a balanced detection capability, with precision and recall exceeding 0.89 for both classes and achieving F1-scores of 0.91 for \textit{Buggy} and 0.9 for \textit{NotBuggy}.

% \begin{table}[ht]

% \caption{Class-wise performance of MLPClassifier.}
% \label{tab:rq3-class-wise}
% \centering
% \begin{tabular}{lccc}
% \toprule
% \textbf{Class} & \textbf{Precision} & \textbf{Recall} & \textbf{F1 Score} \\
% \midrule
% Buggy & 0.90 & 0.91 & 0.91 \\
% NotBuggy & 0.91 & 0.89 & 0.90 \\
% \bottomrule
% \end{tabular}
% \end{table}

% These results indicate that embedding-based ML models, particularly neural networks like MLP, are highly effective for the task of tangled change detection. Their performance not only surpasses other ML models but also rivals that of LLM-based prompt engineering strategies. 

\begin{summarybox}
\textbf{Summary of RQ3}: Embedding-based ML models, particularly MLPClassifier, demonstrate strong predictive power for tangled change detection, achieving over 90\% F1-score and outperforming prior baselines of RQ1 and RQ2. More encouragingly, our results suggest that the performance of embedding-based models can be improved with an enlarged training dataset.  
\end{summarybox}

\subsection{RQ4: Potential impact of LLM-based untangling on future method-level bug prediction models}

From the results of RQ1 to RQ3, we observe that the LLM-based method-level tangled change detection approach performs well with high accuracy and F1-scores using both prompting techniques and embedding-based ML models. This suggests that such approaches can effectively reduce noise in bug datasets. However, an important question remains: \textit{Does this noise reduction benefit machine learning-based bug prediction models?}  Building such a model, however, requires addressing several challenges. First, a broad set of code metrics is needed, whereas the dataset from Chowdhury et al.~\cite{chowdhury_method-level_2024} includes only five. Second, Mashhadi et al.~\cite{mashhadi2023method} demonstrate that combining code metrics with embeddings leads to better performance, requiring us to generate embeddings for all methods. Finally, a thorough evaluation across multiple algorithms and fine-tuning strategies is essential. Due to these requirements, constructing a full bug prediction model is beyond the scope of this paper. Nonetheless, if we can demonstrate that our noise reduction approach improves the ability of code metrics to distinguish between bug-prone and non-bug-prone methods, likely, the performance of machine learning models relying on these metrics will also improve. Therefore, this research question examines whether the distribution of code metrics becomes more distinguishable in the dataset produced by our method.

In RQ1 and RQ2, we evaluated multiple prompting strategies and LLMs, showing that \textit{GPT-4o} combined with few-shot + chain-of-thought prompting delivers the best and balanced performance comparing all metrics. In RQ3, we examined the feasibility of using LLM-generated embeddings with downstream machine learning classifiers to detect tangled changes. Although the latter approach yields better results, it requires generating embeddings for all tangled code diffs from bug-fix commits to support prediction with the machine learning model, which is more complex and time-consuming than the few-shot + chain-of-thought approach. Therefore, we employed \textit{GPT-4o} with few-shot + chain-of-thought prompting to detect tangled changes across all methods in 49 software projects in a fast and efficient manner.

In the dataset by Chowdhury et al.~\cite{chowdhury_method-level_2024}, each method is associated with its change history and the number of other methods modified in the same commit. We apply the few-shot and chain-of-thought strategy using \textit{GPT-4o} to separate \textit{NotBuggy} changes within bug-fix commits, thereby removing noise. To enable replication, the algorithm for creating the Less-Noisy dataset is included in the shared repository (algorithm/algorithm.png). After constructing the \textit{Less-Noisy} dataset, we obtain four types of methods:

\textbf{Noisy NotBuggy}. Following Chowdhury et al.~\cite{chowdhury_method-level_2024}, methods are labeled as \textit{NotBuggy} if they are at least two years old and never involved in a bug-fix. This set is called \textit{Noisy} because it comes from the original noisy dataset, which excludes non-buggy methods that were mixed with buggy changes.

\textbf{Less-Noisy NotBuggy.} It expands on the previous set by including methods confirmed as \textit{NotBuggy} after LLM-based untangling.

\textbf{Noisy Buggy.} Following Chowdhury et al.~\cite{chowdhury_method-level_2024}, methods are labeled as \textit{Buggy} if they appear in bug-fix commits, without accounting for tangling, hence the dataset is considered Noisy.

\textbf{Less-Noisy Buggy.} This set includes methods identified as truly buggy after applying our LLM-based untangling approach.

% \begin{algorithm}[H]
% \small
% \caption{Creation of Less-Noisy Dataset by Untangling.}
% \label{alg:alg1}
% \begin{algorithmic}[1]
% \ForAll{project files}
%     \ForAll{methods}
%         \If{method age $< 2$ years}
%             \State \textbf{continue}
%         \EndIf
%         \If{any commit is buggy (High-Prec.) \textbf{and} tangled}
%             \ForAll{commits}
%                 \If{buggy \textbf{and} tangled}
%                     \State detect with LLM
%                     \If{detection is ``Buggy''}
%                         \State mark \texttt{Buggy}, \textbf{break}
%                     \EndIf
%                 \ElsIf{buggy \textbf{and not} tangled}
%                     \State mark \texttt{Buggy}, \textbf{break}
%                 \EndIf
%             \EndFor
%             \If{none marked \texttt{Buggy}}
%                 \State mark \texttt{NotBuggy}
%             \EndIf
%         \ElsIf{any commit is buggy (High-Prec.)}
%             \State mark \texttt{Buggy}
%         \ElsIf{any commit is \textbf{not} buggy (High-Recall)}
%             \State mark \texttt{NotBuggy}
%         \EndIf
%     \EndFor
% \EndFor
% \end{algorithmic}
% \end{algorithm}

The dataset provided by Chowdhury et al.~\cite{chowdhury_method-level_2024} contains five code metrics computed for each version of a method across its history. For our analysis, we consider the metrics from the first version of each method to ensure consistency across the dataset. These code metrics are defined as follows: \textit{Size} refers to the number of source lines of code, excluding comments and blank lines~\cite{chowdhury_empirical_2022, chowdhury_method-level_2024}; \textit{Readability} is a score that reflects the ease of reading the source code, developed by Buse et al.~\cite{buse2009learning}; \textit{McCabe} measures cyclomatic complexity by counting the number of independent execution paths in a method~\cite{mccabe1976complexity}; \textit{FanOut} represents the number of distinct methods called by a given method, indicating its dependency footprint~\cite{pascarella_performance_2020, chowdhury_method-level_2024}; and \textit{Maintainability Index (MI)} is a composite metric that combines several aspects, including complexity and size, to offer a comprehensive assessment of code maintainability~\cite{oman1992metrics}.

\begin{figure}[ht]
    \centering
    \includegraphics[width=0.96\linewidth]{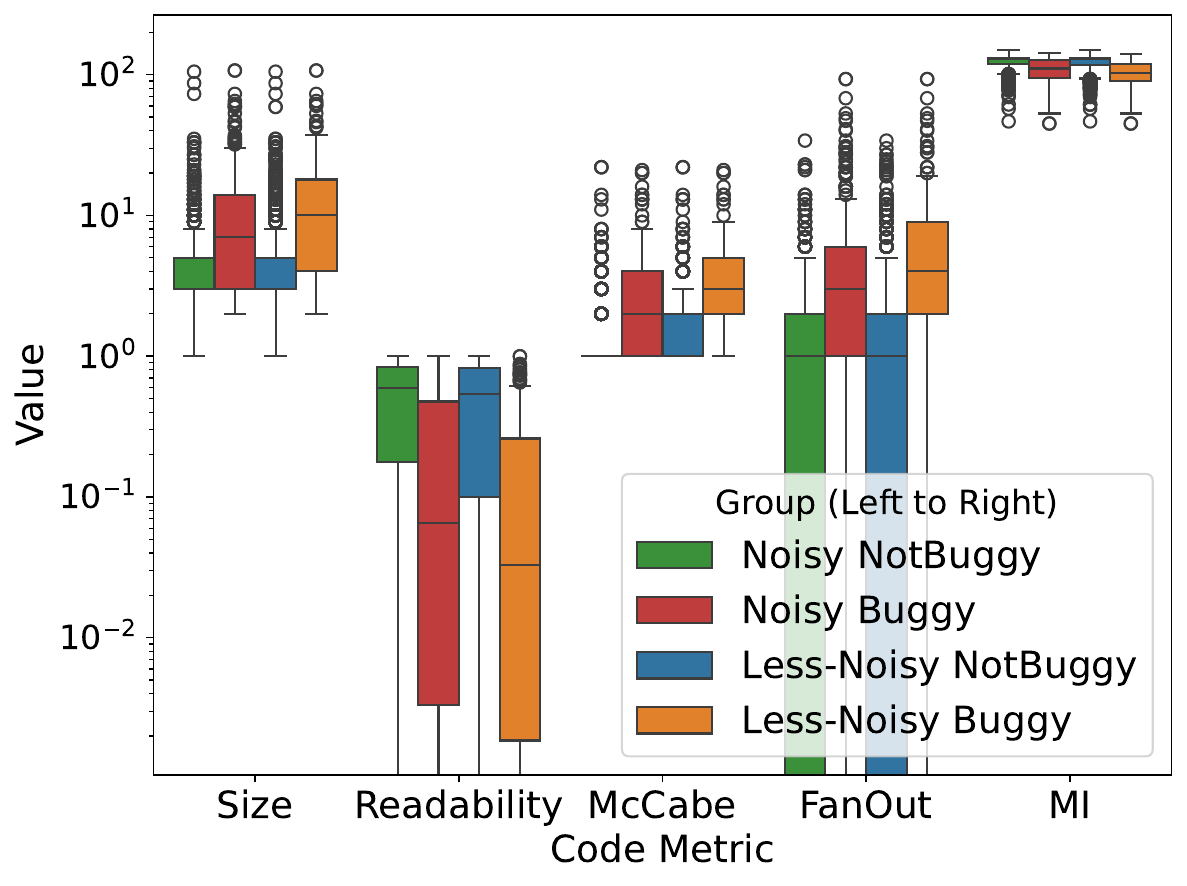}
    \caption{Distribution of code metrics for \textit{Buggy} and \textit{NotBuggy} methods in the \textit{Titan} project across the \textit{Noisy} and \textit{Less-Noisy} datasets. Metric distributions differ more significantly in the \textit{Less-Noisy} dataset.}
    \Description{Boxplot of Titan}
    \label{fig:boxplot}
\end{figure}

Figure~\ref{fig:boxplot} presents the distribution of these 5 code metrics across the four method sets for the project \textit{Titan}. Visually, the difference in \textit{Size} between \textit{Less-Noisy Buggy} and \textit{Less-Noisy NotBuggy} methods appears larger than the other two sets. A similar pattern is observed for all other code metrics. To determine whether these differences are statistically significant, we use the Wilcoxon rank-sum test~\cite{mcknight2010mann}. This non-parametric test is widely adopted in software engineering research due to its robustness in comparing distributions~\cite{bangash2020time, chen2020savior, chowdhury2019greenscaler}. The results show that the differences in distributions are indeed statistically significant ($p \le 0.05$) for all code metrics. 

As the differences are statistically significant for both datasets, we calculated Cliff's delta effect size~\cite{macbeth2011cliff} to understand the magnitude of these differences. This non-parametric measure is also widely used in software engineering research~\cite{bangash2020time, chen2020savior, chowdhury2019greenscaler} and does not require assumptions about the data distribution. The results show that, in the \textit{Noisy} dataset, the effect size is large for \textit{MI} and medium for all other code metrics. In the \textit{Less-Noisy} dataset, the effect size is large across all code metrics.

To evaluate the generalizability of our results, we performed Wilcoxon rank-sum tests across all 49 projects. In the \textit{Noisy} dataset, significant differences ($p \le 0.05$) were observed in 47 projects (95.92\%) for \textit{Size}, \textit{McCabe}, and \textit{MI}, 45 projects (91.84\%) for \textit{Readability}, and 48 projects (97.96\%) for \textit{FanOut}. In the \textit{Less-Noisy} dataset, we observe improved results, with one additional project exhibiting statistically significant separation for each metric.

To further quantify the differences between \textit{Buggy} and \textit{NotBuggy} methods, we compute Cliff’s Delta effect sizes across all 49 projects. Table~\ref{tab:cd_both} presents the effect size distributions for both the \textit{Noisy} and \textit{Less-Noisy} datasets. We categorize the effect sizes following Hess et al.~\cite{hess2004robust}: \textit{Negligible} ($< 0.147$), \textit{Small} ($0.147 \leq \delta < 0.33$), \textit{Medium} ($0.33 \leq \delta < 0.474$), and \textit{Large} ($\delta \geq 0.474$).

\begin{table}[h]
\centering
\caption{Cliff's Delta effect size values compare \textit{Noisy} and \textit{Less-Noisy} datasets, grouped by effect size category and shown as percentages in the format \textit{Noisy / Less-Noisy}. The percentages for the \textit{Noisy} dataset do not sum to 100\% because the \textit{RxJava} project was excluded as it had no \textit{NotBuggy} methods.}
\resizebox{\columnwidth}{!}{%
\addtolength{\tabcolsep}{-0.15em}
\begin{tabular}{lcccc}
\toprule
\textbf{Metric} & \textbf{Negligible} & \textbf{Small} & \textbf{Medium} & \textbf{Large} \\
\midrule
Size        & 0 / 0     & 14.29 / 6.12     & 46.94 / 20.41    & 36.73 / 73.47    \\
Readability & 2.04 / 2.04     & 36.73 / 24.49    & 42.86 / 40.82    & 16.33 / 32.65    \\
McCabe      & 2.04 / 2.04     & 34.69 / 8.16     & 36.73 / 42.86    & 24.49 / 46.94    \\
FanOut      & 0 / 0     & 10.20 / 6.12     & 46.94 / 22.45    & 40.82 / 71.43    \\
MI          & 0 / 0     & 12.24 / 6.12     & 34.69 / 16.33    & 51.02 / 77.55    \\
\bottomrule
\end{tabular}%
}
\label{tab:cd_both}
\end{table}

The table reports the proportion of projects falling into each effect size category for all five code metrics. The results indicate that, for most metrics, the \textit{Less-Noisy} dataset shows a substantial shift toward larger effect sizes. For example, for the code metric, \textit{Size}, in the \textit{Less-Noisy} dataset, 73\% of the projects exhibit a large effect size in the difference between \textit{Buggy} and \textit{NotBuggy} methods, compared to only 36\% in the \textit{Noisy} dataset. Similar patterns are observed for the other four code metrics. These findings clearly demonstrate that our filtering approach effectively eliminates noise and increases the distinction between the code metrics of buggy and non-buggy methods. As a result of this enhanced separability, we believe that our approach will support improved performance in future machine learning models for bug prediction.

\begin{summarybox}
\textbf{Summary of RQ4:} After constructing the \textit{Less-Noisy} dataset using the LLM-based approach, statistical analysis reveals significantly improved separability (with larger effect sizes) in code metrics between buggy and non-buggy methods. This suggests that the noise reduction could enhance the performance of future ML-based bug prediction models.
\end{summarybox}

\section{Threats to Validity}
\label{sec:threats}
\textit{Construct Validity.} The  manual annotation of code diffs as \textit{Buggy} or \textit{NotBuggy} may be affected by subjective bias. Although annotators were provided with explicit guidelines and a UI tool to assist in their evaluations, subjective judgment and semantic ambiguity in commit messages may have influenced labeling outcomes.

\textit{External Validity.} The dataset used in this study comprises only open-source Java projects, which may limit the generalizability of our findings. Applying this approach to other programming languages, project domains, or commit conventions may require modifications to the prompts or retraining of models.

\textit{Internal Validity.} The dataset used in this study was curated from method histories of Java projects obtained through CodeShovel~\cite{grund_codeshovel_2021}, which is not fully accurate. Any inaccuracies in the method histories could impact the results.

\textit{Conclusion Validity.} It is affected by all of the above mentioned threats.

\section{Conclusion and Future Work}
\label{sec:conclusion}
Method-level bug prediction is considered one of the holy grails in software engineering research. However, it remains an open challenge, primarily due to the absence of a noise-free method-level bug dataset~\cite{chowdhury_method-level_2024, pascarella_performance_2020}. To address this, we evaluated the effectiveness of various LLMs and prompting strategies for detecting tangled changes at the method level. Our results show that even zero-shot LLMs achieve high accuracy when both commit messages and code diffs are provided (RQ1). Performance improves further when combining chain-of-thought with few-shot prompting (RQ2). Additionally, embedding-based models deliver even greater accuracy gains (RQ3). Across RQ1 to RQ3, open-source models also demonstrate strong performance. While they do not yet match proprietary models, they offer efficient alternatives with acceptable trade-offs. Building on these insights, we generated a \textit{Less-Noisy} dataset using our LLM-based method. This dataset shows promise in developing more accurate method-level bug prediction models as it exhibits a significantly stronger power of commonly used code metrics to differentiate between bug-prone and non-bug-prone methods compared to the original \textit{Noisy} dataset (RQ4).

The implications of our findings extend beyond the untangling problem and contribute to broader discussions on the capabilities and limitations of LLMs in software engineering. Although challenges such as hallucinations, reproducibility, and data leakage remain~\cite{yao_llm_2024, ji_survey_2023, ye_cognitive_2023, sallou_breaking_2024}, our study offers empirical evidence supporting the effectiveness of LLMs in tasks such as code reasoning, classification, and dataset construction~\cite{zheng2024few, jaoua_combining_2025, nam_using_2024}. In particular, our work shows that when guided by carefully designed prompts, LLMs can significantly improve software engineering workflows.

Looking ahead, a promising direction for future research is to enrich the dataset with additional code metrics and embeddings and to explore a range of machine learning models for building more effective method-level bug prediction models. To support continued progress, we have publicly released our code and dataset, allowing researchers to use them directly or adapt them to their specific needs.

\bibliographystyle{ACM-Reference-Format}
\bibliography{sample-base, new_references}

%\appendix

\end{document}